# The near surface vertical atmospheric electric field abnormality could be as a promising imminent precursor of major earthquakes

T. Chen[1],[*], H. Wu[1,2], X.-X. Zhang[3], C. Wang[1], X.-B. Jin[4], Q.-M. Ma[5], J.-Y. Xu[1], S.-P. Duan[1], Z.-H. He[1], H. Li[1], S.-G. Xiao[1], X.-Z. Wang[6], X.-H Shen[7], Q. Guo[7], I. Roth[8], V. S. Makhmutov[9], Y. Liu[1], J. Luo[1], X.-J. Jiang[1], L. Dai[1], X.-D. Peng[1], X. Hu[1], L. Li[1,2], C. Zeng[1,2], J.-J. Song[5], F. Xiao[5], J.-G. Guo[3], C. Wang[3], H.-Y. Cui[10], C. Li[10], Q. Sun[11]

[1] State Key Laboratory of Space Weather, National Space Science Center, Chinese Academy of Sciences, Beijing, China.

[2] University of Chinese Academy of Sciences, Beijing, China.

[3] National Center for Space Weather, Chinese Meteorological Administration, Beijing, China.

[4] Sichuan Meteorology Agency, Chinese Meteorological Administration, Chengdu, China.

[5] Institute of Electrical Engineering, Chinese Academy of Sciences, Beijing, China.

[6] Institute of geophysics, Chinese Earthquake administration, Beijing, China.

[7] Institute of Crustal Dynamics, Chinese Earthquake administration, Beijing, China.

[8] Space Science Laboratory, UC Berkeley, CA, USA.

[9] Lebedev Physical Institute, RAS, Moscow, Russia.

[10] Institute of Acoustics, Chinese Academy of Sciences, Beijing, China.

[11] Institute of Atmospheric Physics, Chinese Academy of Sciences, Beijing, China.

*Corresponding author. Email: tchen@nssc.ac.cn

**Abstract:** A promising short term precursor of major earthquakes (EQ) is very crucial in saving people and preventing huge losses. $E_z$, atmospheric electrostatic field vertical component, under fair air conditions, is generally oriented downwards (positive). Anomalous negative $E_z$ signals could be used as an indicator of a great number of radioactive gases which are released from great number of rock clefts just before major earthquakes. Enhanced emission of radon radioactive decay will produce an anomalously large number of ion pairs. The positive particles will be transported downward by the fair weather electrostatic field and pile up near the surface. Finally, obviously and abnormally, an oriented upward atmospheric electric field $E_z$ near the ground could be formed. Therefore, monitoring this $E_z$ may be applied effectively in earthquake warning.

Due to the devastating consequences of major earthquakes, the identification of any reliable precursor signatures is of paramount importance. There are many kinds of parameter anomalies which could be applied as the earthquake precursors (*1*): geostress change; underground water level and other fluid change; geoelectric field, geo-conductivity and geomagnetic disturbance record; gravitational anomalies; surface deformation , huge gas fluxes out of the crust; variation of radon gas; temperature variations of the Earth's surface; air temperature variations; variations of air relative humidity; anomalous flux of latent heat of evaporation; extraordinary vertical profiles of air temperature and humidity; linear cloud anomalies; anomaly of radio waves propagation in VLF, HF and VHF frequency bands;

extraordinary concentration and distribution of aerosols; anomalies of the outgoing longwave radiation OLR energy flux; local (in situ) anomalies of space plasma parameters (concentration of ions and electrons, ion and electron temperature, mass composition and concentration of the major ions); extraordinary ELF and VLF emission measured on board the satellite, quasi-constant magnetic and electric fields; extraordinary particle precipitation fluxes for different energy bands; vertical profiles electron concentration; extraordinary TEC change by GPS data processing (*1*). Additionally, atmospheric electrostatic field monitoring becomes more appealing these days. The atmospheric electric field anomaly prior the earthquake has ever been widely studied (*2-6*). Since the Tangshan earthquake in 1976, China has set up several monitoring stations of atmospheric electric fields, and observed some obvious anomalous cases (*7*). Yashutaka et al., (*8*) had pointed out that the anomalous radon emission triggers a great increase in the number density of small (or light) ions and in the atmospheric conductivity and the decrease in the atmospheric field of the lower atmosphere (from the ground to the altitude of 2 km) as observed around the time of the Kobe earthquake in 1995. Yashutaka et al., (*8*) had suggested further that the behavior of radon in terms of the atmospheric electrical quasi-static process can explain seismic precursors observed near the ground. Choudhury et al. (*2*) had described the characteristics of the vertical atmospheric electrostatic field being negative 7-12 hours just before earthquake according to the statistics of 30 various class earthquake events over northern India. Altogether, there exists a unique feature that the low altitude atmospheric quasi-static electric field appears anomalously negative just before some earthquake events with fair weather condition (*7, 9*).

Although the observation of atmospheric electric field before the earthquake was carried out earlier, it requires additional investigation. On one hand, it is difficult to extract the earthquake precursor information in the atmospheric electric field anomaly; intense convection due to weather patterns, space weather effects, human activities, etc. may strongly interfere with the atmospheric ion background to produce ionization and change motion of charged ions which are coupled to neutral molecules. On the other hand, there exists a lack of sufficient, credible, and large-amount electric field data for researchers to study and improve the prediction of earthquakes by paying attention to atmospheric electric field negative anomalies (*3*). Also, until now, no compelling physical mechanism exists which can explain why an abnormal atmospheric electrostatic field reverse signal could appear just before an earthquake. In this paper, we emphasize the importance of the negative anomalous signals of the atmospheric electric field to help predicting some forthcoming earthquakes. First, we present a set of electric field measurements at several sites over period of hours prior to the occurrence of earthquakes. Then, we offer an explanation of the electric field anomaly as a result of the enhanced production of ionization radiation due to tectonic activity preceding the earthquake. Finally, it is pointed that monitoring the near ground atmospheric electrostatic field $E_z$ anomalous negative signal may be very useful in major earthquake forthcoming warning.

**$E_z$ observation results just before some earthquake events**

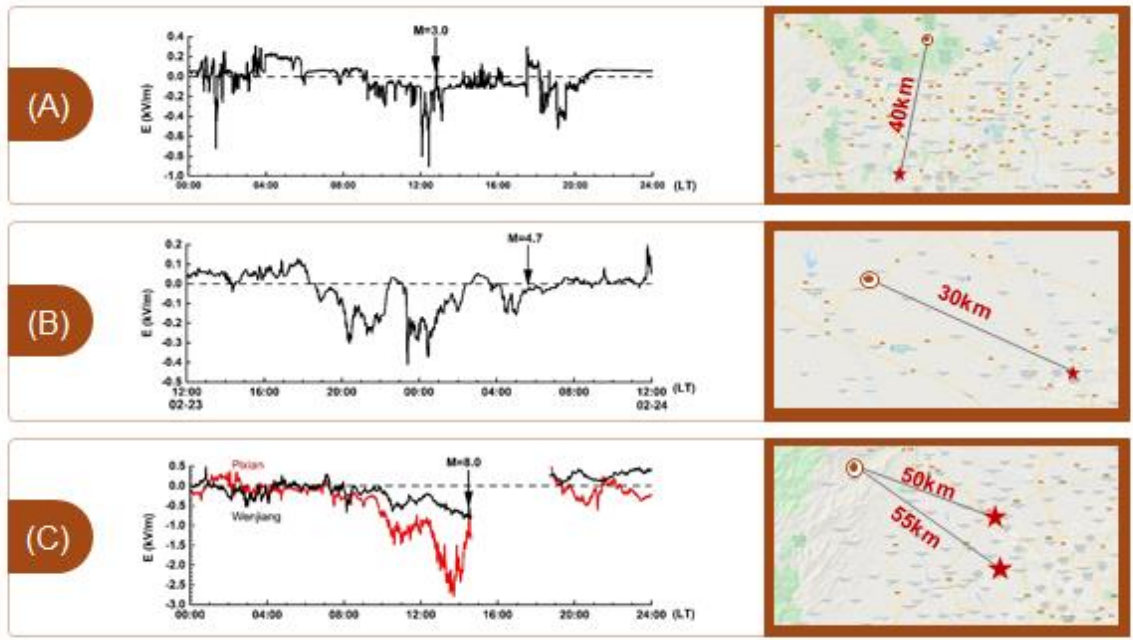

**Fig. 1. Negative atmospheric electrostatic field Ez before earthquakes.** (**A**) Beijing M3.0 earthquake on 14 April, 2019, with the station 40 km from the epicenter. (**B**) Rongxian M4.7 earthquake on 24 February, 2019, with the station 30 km from the epicenter. (**C**) Wenchuan M8.0 earthquake on 12 May, 2008, with two different stations 50 and 55 km from the epicenter. The data gap at the bottom panel 3 is due to power off. On the right maps, the stars note the stations, the oval notes the epicenter.

On 14 April of 2019 in Beijing, China, the weather has been identified to be fair based on the criteria proposed by Harrison and Nicoll (*10*). An abnormal phenomenon was observed when the normal positive $E_z$ (+100 V/m, pointing downward, measured on the top of a building) changed into negative (about -100 V/m) around 09:00 LT in the morning, as shown in Fig. 1A. It had been predicted by the corresponding author that a medium magnitude earthquake will occur several hours later based on previous analytic results (*3, 5, 11*). Later, a M3.0 earthquake was confirmed to occur at 12:47 LT with the epicenter 40 km north of the place where the scientific instrument was located.

Similarly, as shown in Fig. 1B, when the Rongxian M4.7 earthquake occurred at 05:38:10 LT on February 24, 2019 (Lon=104.49°, Lat=29.47°) in Sichuan Province, while at Zigong station situated 30 km from the epicenter, under fair air conditions the $E_z$ became negative about 11 hours earlier. The $E_z$ was suddenly reduced to a minimum of -410 V/m at 23:24 LT (maximum of the negative $E_z$). Although there are two short intervals of positive $E_z$ (tens of minutes), the Ez had shown negative anomaly for more than ten hours until the earthquake began.

Another example is shown in Fig. 1C. In spite of the data gap, in Pixian and Wenjiang counties Ez also acquired negative values for 7 hours before the Wenchuan M8.2 earthquake which occurred at 14:28 LT on May 12 of 2008. Note that the weather in Wenchuan, Pixian and Wenjiang counties all day was very fair as well. The peak negative value of $E_z$ at Pixian and Wenjiang station is -2750 V/m and -750 V/m, respectively. The Pixian station is 50 km off the epicenter, while the Wenjiang station is 55 km off the epicenter.

All the above examples indicate that there exists a negative atmospheric electrostatic field anomaly several hours before the occurrence of some earthquakes. It is demonstrated that if the atmospheric electric field $E_z$ near the earth surface shows a hourly stable negative value (-100 V/m ~ -3000 V/m) under fair air conditions, meaning that without air pollution or sandy wind, an earthquake (M:3-8) may be likely to take place in very short term (sometimes up to ten hours) within a distance of less than 100 km. The greater the intensity of the earthquake, the more easily the reversal abnormality of the $E_z$ signal could be identified that might be applied for earthquake prediction in the near future.

**Discussion**

Why does the atmospheric electrostatic field $E_z$ reversal last for several hours just before some earthquakes? It could be explained that before the fault elastic rebound, the crust movement may make huge number of rock clefts and releases significant amount of radioactive radon gas into the atmosphere near the Earth's surface by these clefts. The radon gases further decay and emit α particles, β particles and γ rays. Compared to β particle and γ rays at various altitudes, α particle has more powerful ionizing radiation capacity in the air (*12*). Therefore, these α particles are the main source for ion pairs that include the positive and negative charged particles. In turn, no matter at how high altitude they are produced, these positive charged particles could be driven downward to the earth by the fair weather Ez, and gradually piled up above the surface. Further, these positive charged particles distribution form a vertical gradient in charge density. This gradient will produce the "reverse" electric field. The magnitude of the "reverse" electric field may exceed that of the formal downward Ez in the fair air (shown as Fig. 2D).

Under fair weather conditions, appearance of the upward electric field $E_z$ before major earthquake indicates that the daily normal positive atmospheric electric field $E_z$ and the part induced by space weather (*13*) had been overcome, so, $E_z$ presents some negative signatures, which may last for several hours or more. Radon gas ionizing the air after its escape from the rocks due to strong tectonic activity near the epicenter region has been suggested by previous researchers (*14-21*). The physical process that produces hour-scale persistent negative anomaly of atmospheric electrostatic field near the epicenter before earthquake could be illustrated in Fig. 2.

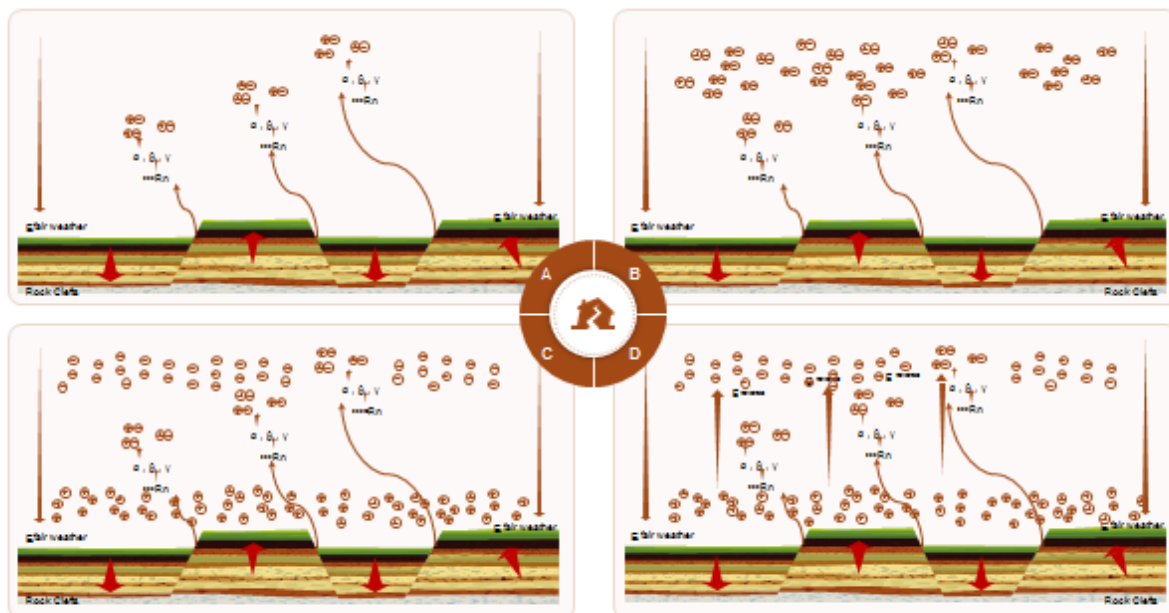

**Fig. 2. Schematic diagram of the seismogenic process changing Ez.** (**A**) In fair weather background, normal vertical atmospheric electric field Ez orients downward (the downward direction of Ez is defined as positive under fair weather condition). The seismogenic process releases radon from rock fracture and produces α particle decay to complete ionizing radiation, so many positive and negative charged particles are injected close to the surface near the epicenter. (**B**) Greater numbers of electron-ion pairs disperse in the air above the local region near the epicenter. (**C**) Positive ions and negative ions which bring different charges will depart by the fair electric static field Ez, go down and lift up respectively. (d) The earthquake related Ez (upward direction: negative) represent the electric field generated by a series of physical processes that radon ionizes neutral gas in the air.

Fig. 2 has demonstrated the physical mechanism that the seismogenic process changes $E_z$ in direction and magnitude. Fig. 2A: In fair weather background, normal vertical atmospheric electric field $E_z$ orients downward (the downward direction of $E_z$ is defined as positive under fair weather condition,). The seismogenic process always releases radon from many rock clefts. In turn, radon could generates radioactive decay and emits α, β and γ. These rays may further ionize the air nearby. With more and more emission of radon gases, so many positive and negative charged particles are injected close to the surface near the epicenter. Fig. 2B: Greater numbers of electron-ion pairs disperse in the air above the local region near the epicenter. Fig. 2C: Positive ions and negative ions which bring positive charges and negative charges respectively will depart by the fair electric static field $E_z$, go down and lift up separately. Fig.2D: The earthquake related $E_z$ (upward direction: negative) represent the electric field generated by a series of physical processes that radon ionizes neutral gas in the air. It illustrates that in the last stage of preearthquake, largely release of radon gas near the imminent fracture region will finally lead to the vertical atmospheric electrostatic field $E_z$ shows its unique negative (upward) orientation.

As shown as Fig.2, the positive particles which are emitted at various altitudes will always be driven downward by the fair weather $E_z$. So, gradually piled positive particles will produce an $E_z$ upward. In other word, a "reverse" $E_z$ is formed near the earth surface. Finally, a very strong and stable electrostatic field has been established by positively charged particles and show Ez negative. In general, the phenomenon lasts a few hours just before the earthquake.

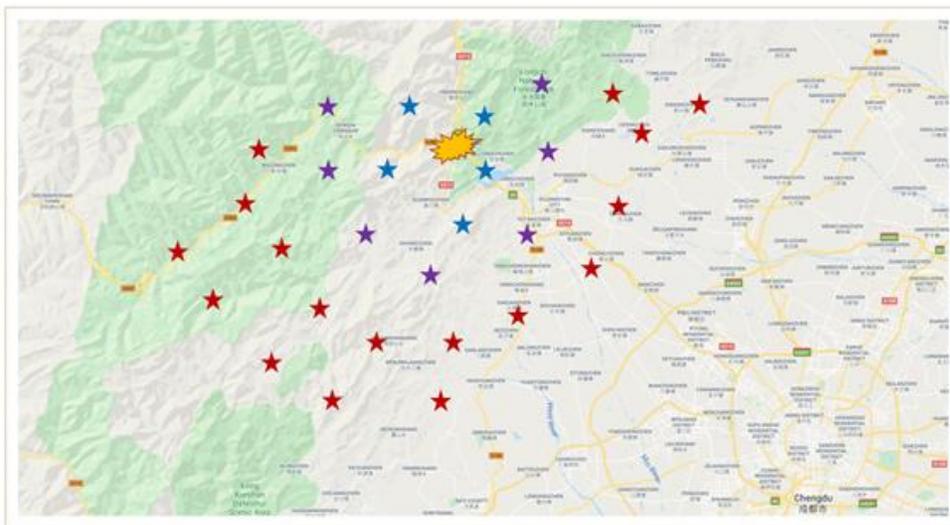

**Fig. 3. Cartoon map of monitoring network for waning major earthquakes**. Star notes station. For 5 class EQ, Ez anomalous negative signals may be recorded by blue stations. For 6 class EQ, Ez anomalous negative signals may be recorded by blue and purple stations. For 7 class EQ, Ez anomalous negative signals may be recorded by blue, purple and red stations. Yellow oval note the epicenter.

Fig. 3 is a cartoon map which indicates that a local monitoring network could be useful in warning major earthquakes.

Fig. 1C has also illustrated that a pre-earthquake electrostatic field which appeared in Wenchuan region and nearby just before the M8.0 earthquake. The earthquake related reverse electrostatic field may be inhomogeneous in the whole fault region. The closer to the epicenter, the greater the negative $E_z$ value is measured by the instrument. Therefore, before Wenchuan earthquake, the value of the $-E_z$ measured by Pixian station (50 km off the epicenter) is greater (4 times) than that of Wenjiang station (55 km off the epicenter), while both the two stations simultaneously showed that $-E_z$ reversal appeared from 7:00 LT till the earthquake occurrence. The total $E_z$ negative anomaly lasted 7 hours (shown as Fig. 1C). The comparison between the two simultaneous $E_z$ values from different observation sites demonstrates that the closer the $E_z$ monitoring station is to the epicenter, the greater the magnitude of the earthquake related reverse $E_z$ signal is obtained by the monitoring site. Based on the above observations, monitoring stations that may observe the reverse $E_z$ signal should be off the epicenter in the distance less than 100 km. If there are many nearby stations, someone could determine the location of the epicenter by Ez data from multiple stations and triangulations.

It is suggested that with the establishment of more stations to monitor the $E_z$, one can determine more accurately what magnitude, when, and where the major earthquake will occur. Geological conditions, surface structure in epicenter region, wind speed and wind velocity direction near the station during observation will affect the magnitude and lasting time of the negative $E_z$ value related to EQ. Therefore, the selection of the observed sites should be established according to the localized patterns of the fault zones in the future.

The stronger an earthquake is, the greater the signal magnitude of the $E_z$ negative value may be. In general, the magnitude of a potential earthquake could be proportional to the area of the fault, the number and the depth of the rock clefts due to drastically tectonic cracks in the last stage of the pre-earthquake. It can calibrate some warning parameters based on the negative magnitude of the Ez according to localized geological conditions. If in fair condition, an instrument perceives the stable negative $E_z$ value, the forthcoming earthquake might occur nearby, in general, in a distance of less than 100 km.

Exampled by the cartoon map shown in Fig. 3, for a 5 class earthquake, it may cause several stations noted by blue color of near the epicenter to receive negative anomalous signals, and for a 6 class earthquake, it may cause more than ten stations (noted by blue color and purple color) nearby the epicenter to receive negative anomalous signals, and for a 7 class earthquake, it may cause more than twenty stations (noted by blue color, purple color and red color) nearby the epicenter to receive negative anomalous signals.

Now, we can summarize and answer the scientific question: "Why could near surface vertical atmospheric electrostatic field always be negative just before earthquake" with the following descriptions.

During final stage, drastic changes in fault movement just before major earthquakes result in the following;

1). More radon gases will be released from rock clefts.

2). Radon radioactive decay processes in the air will produce more α, β and, γ radiation.

3). More ion pairs (positive and negative ions) will be born in the processes due to α particle's ionization radiation.

4). In the meantime, if some major earthquake could be forthcoming, latent heat increase, the weather must change into quasi-fair weather which makes someone easy to distinguish reverse Ez anomalies (*1, 18*).

5). Positive ions could be transported to surface by the downward $E_z$ in fair weather condition.

6). More positive charges will be piled up near the surface and make $E_z$ reverse near the surface (shown as Fig. 2D).

7). The greater the class of the EQ is, the easier the $E_z$ anomalous signal could be recognized.

Therefore, $E_z$ being negative anomalous could be used as an important imminent precursor of major earthquakes. Fig. 4 has illustrated the understanding to the real physical process and future warning application.

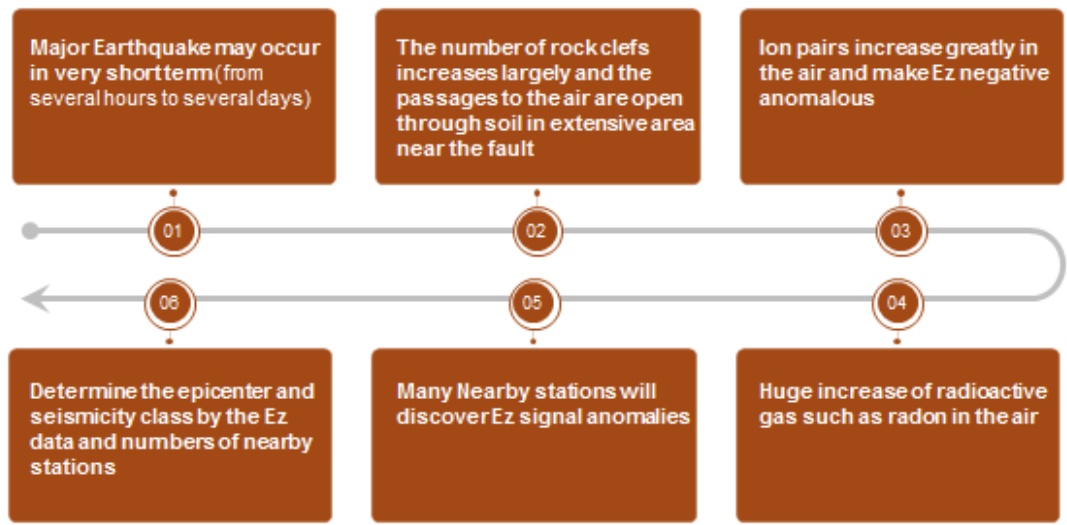

**Fig.4. The application process of the near surface Ez anomalous signals as a promising imminent precursor of major earthquakes.**

**Acknowledgments:** The authors thank Prof. Fushan Luo, Prof. Jie Liu, Prof. Zhijun Niu and Prof. Shi Che for very useful discussion. **Funding:** Supported by the Strategic Pioneer Program on Space Science, Chinese Academy of Sciences, Grant No. XDA17010301, XDA17040505, XDA15052500, XDA15350201, and by the National Natural Science Foundation of China, Grant No. 41731070, 41931073. The authors thank some supports from the Specialized Research Fund for State Key Laboratories, and CAS-NSSC-135 project. **Author contributions:** Tao Chen designed the project. Han Wu, Xiaoxin Zhang, Xiaobing Jin, Lei Li and Jianguang Guo, Chong Wang collected the atmospheric electrostatic field Ez data and analyzed them compared with the earthquake data. Professor Chi Wang, Jiyao Xu, Hui Li, Yong Liu, Suping Duan, Saiguan Xiao, Lei Dai, Ilan Roth, Vladimir Makhumutov and Xiong Hu contribute space physics, atmospheric physics, atom physics theory and signal analysis. Qiming Ma, Jiajun Song, Fang Xiao, Jing Luo, Zhaohai He, Chen Zeng, and Xujie Jiang are responsible for building the atmospheric electric field instrument and calibration work and the Ez measurement. Xichen Wang, Xuhui Shen, Quan Guo, Xiaodong Peng, Hanyin Cui, Chao Li, Qiang Sun provide sub-sonic and earthquake data and do some relative data analysis. Tao Chen wrote the paper with input from all authors. **Competing interests:** The authors declare that there are no competing interests. **Data and materials availability:** all data used in the research article are from Chinese Meorological Administration, Chinese Earthquake administration and Chinese Academy of Sciences.

# Supplementary Materials

**Materials and Methods**

Chinese meridian Project supervised by Chinese Academy of Sciences and many weather monitoring stations supervised by China Meteorology Administration have offer much date from the instrument that can measure the vertical atmospheric electrostatic field. The accurate of the data is 10 V/m. The normal diurnal curve of atmospheric electrostatic field is called Carnegie curve (*22*), with a typical positive $E_z$ from tens of V/m to two hundred of V/m. Due to the variability of solar activity, $E_z$ could be enhanced to even 800 V/m (*13*).

In order to exclude some obvious weather signals, we need to observe the data of atmospheric electric field under fair weather conditions. The fair weather conditions (*10*) which are used here for deleting any meteorological disturbance that may produce Ez negative signal are as follows: (1) Lower relative humidity (No charged particles influenced by rain, snow or fog; less floating dust, snow; absence of aerosol and haze; visibility is greater than 2 km). (2) No obvious cumuliform and no plenty of stratus clouds with cloud base below 1.5 km. (3) The wind speed should less than 8 m/s within 10 meters of the surface.

Fig. S1 displays the observations of atmospheric electric field under fair weather conditions. All the stations show a positive atmospheric electric field in a whole day under fair weather conditions. If a meteorological instrument is located near the electric field observation instrument, we can judge the cause of the negative anomaly to a certain extent. We have tested the negative anomaly observed in one week. Fig. S2 show three cases of negative anomaly of atmospheric electric field. On the day of the earthquake (Shown in Fig. S2A), we observe that relatively high temperature, low relative humidity (before earthquake occurred at 12:47). And despite the higher wind speed, but lower $PM_{10}$ index indicated that no significant dust that may affected the electric field during that day. Also, the stable positive signal between 7 o'clock and 9 o'clock before the earthquake indicates that the electric field value is not particularly disturbed by meteorological factors and human activities.

Compare with that, a very high relative humidity (Average value is greater than 80 %) can be observed on April 20, which indicated that the negative anomaly is due to the clouds or some rains. On April 21, the most characteristic is the electric field signal changed rapidly. During that day, high value of $PM_{10}$ index was observed. And based on the statistical studies of electric field signal in Beijing by Wu et al. (*23*), the negative rapidly-changed value of atmospheric electric field can be explained as caused by sand particles and air pollutants.

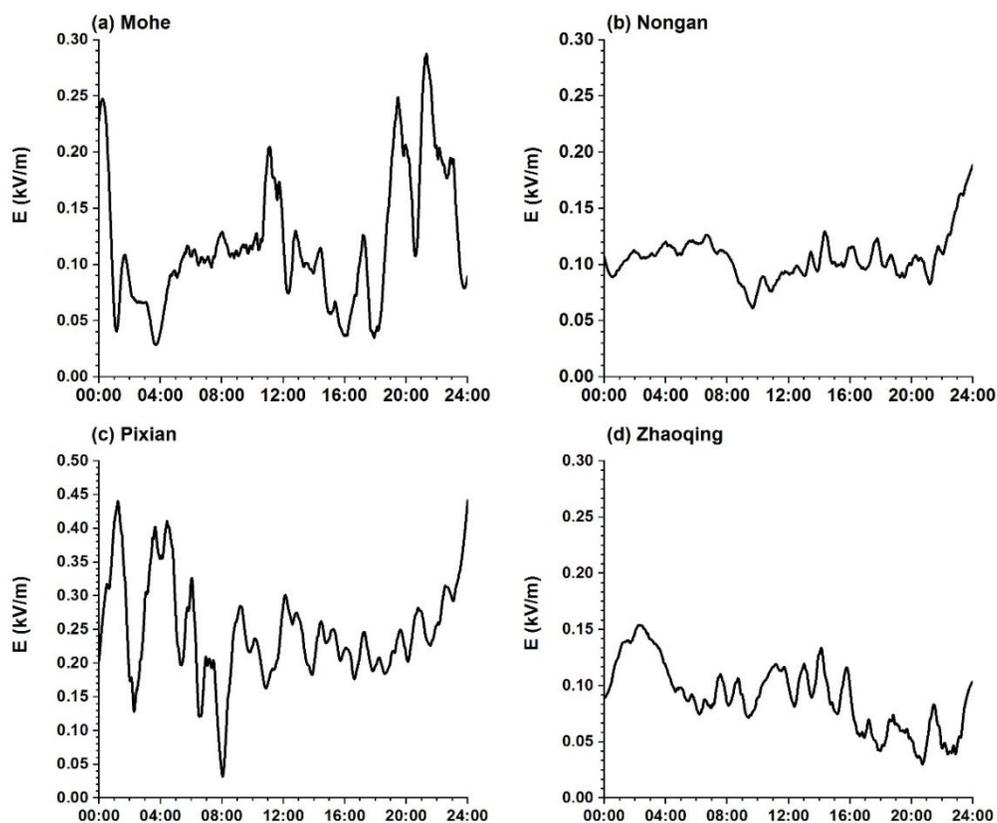

**Fig. S1. One-hour moving average of the atmospheric electric field in different stations of China under fair weather conditions.**

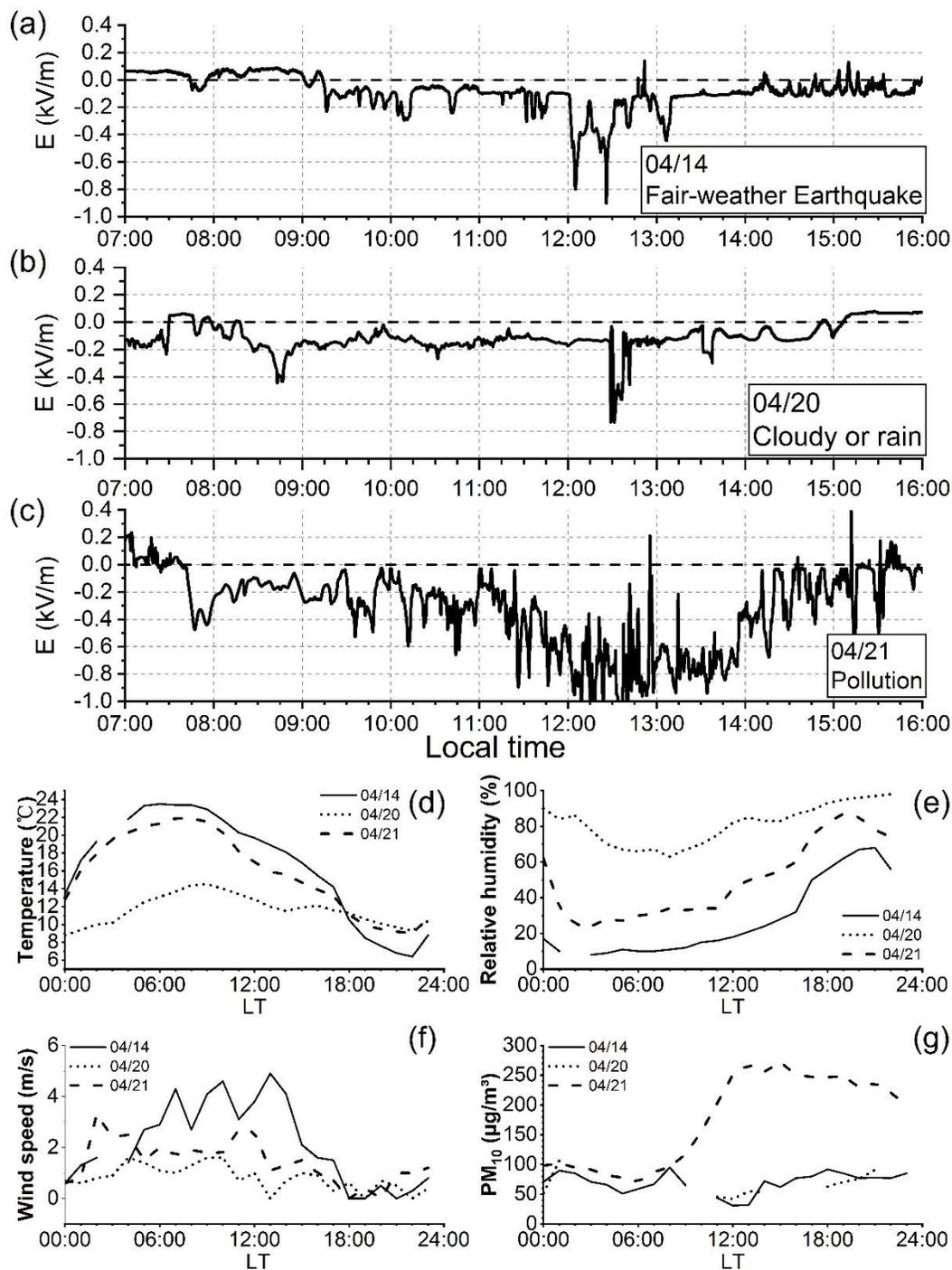

**Fig. S2. The comparison of coming seismology precursor characteristics with other weather negative $E_z$ signal characteristics in 2019.** The meteorological observatories (provide temperature, wind speed, relative humidity elements) and pollution index monitoring instruments (provide $PM_{10}$ data) are within 5 km of the electric field instrument.